\documentclass[preprint,12pt]{elsarticle}
\makeatletter
\def\ps@pprintTitle{%
 \let\@oddhead\@empty
 \let\@evenhead\@empty
 \def\@oddfoot{\textit{Submitted Preprint} \hfill}%
 \let\@evenfoot\@oddfoot}
\makeatother

\usepackage{amssymb}
\usepackage{amsmath}
\usepackage{amsthm}

\usepackage{amssymb}
\usepackage{amsmath}
\usepackage{color,soul}
\usepackage{bbm}
\usepackage{graphicx}     
\usepackage{nicematrix}
\usepackage{natbib}
\usepackage{adjustbox}
\usepackage{multirow}
\usepackage{array}
\usepackage{booktabs}
\usepackage{mathtools}
\usepackage{multicol}
\usepackage{algpseudocode}
\usepackage{algorithm}
\usepackage{xcolor}
\usepackage{caption}

\newtheoremstyle{withoutaperiod}% name
  {0pt}% space above
  {0pt}% space below
  {}% body font
  {}% indent amount
  {\bfseries}% theorem head font
  {}% punctuation after theorem head
  {0.5em}% space after theorem head
  {}% theorem head spec
\newtheorem*{pf}{Proof}
\theoremstyle{withoutaperiod}
\newtheorem{thm}{Theorem}
\newtheorem{cor}{Corollary}

\begin{document}

\begin{frontmatter}



\title{Reinforcement Learning-based Control via Y-wise Affine Neural Networks (YANNs)}


\author[WVU]{Austin Braniff}\ead{austin.braniff@mail.wvu.edu}
\author[WVU]{Yuhe Tian\corref{cor}}\ead{yuhe.tian@mail.wvu.edu}

\affiliation[WVU]{organization={Department of Chemical and Biomedical Engineering, West Virginia University},%Department and Organization
            city={Morgantown},
            state={West Virginia},
            country={United States}}
\cortext[cor]{Corresponding author}

%% Abstract
\begin{abstract}
This work presents a novel reinforcement learning (RL) algorithm based on Y-wise Affine Neural Networks (YANNs). YANNs provide an interpretable neural network which can exactly represent known piecewise affine functions of arbitrary input and output dimensions defined on any amount of polytopic subdomains. One representative application of YANNs is to reformulate explicit solutions of multi-parametric linear model predictive control. Built on this, we propose the use of YANNs to initialize RL actor and critic networks, which enables the resulting YANN-RL control algorithm to start with the confidence of linear optimal control. The YANN-actor is initialized by representing the multi-parametric control solutions obtained via offline computation using an approximated linear system model. The YANN-critic represents the explicit form of the state-action value function for the linear system and the reward function as the objective in an optimal control problem (OCP). Additional network layers are injected to extend YANNs for nonlinear expressions, which can be trained online by directly interacting with the true complex nonlinear system. In this way, both the policy and state-value functions exactly represent a linear OCP initially and are able to eventually learn the solution of a general nonlinear OCP. Continuous policy improvement is also implemented to provide heuristic confidence that the linear OCP solution serves as an effective lower bound to the performance of RL policy. The YANN-RL algorithm is demonstrated on a clipped pendulum and a safety-critical chemical-reactive system. Our results show that YANN-RL significantly outperforms the modern RL algorithm using deep deterministic policy gradient, especially when considering safety constraints.
\end{abstract}

%% Keywords

\end{frontmatter}


\section{Introduction}
Reinforcement learning (RL) has emerged as one of the most promising technologies in the modern era \cite{dogruReinforcementLearningProcess2024,shinReinforcementLearningOverview2019a}. There has been a significant surge in the research focused on this problem-solving strategy, since the seminal work showing the ability to provide human-like control for video game playing by integrating neural networks (NNs) into the RL algorithms \cite{mnihHumanlevelControlDeep2015c}. RL has shown great promise in many areas including game playing \cite{silverMasteringGameGo2017,silverMasteringGameGo2016}, robotics \cite{kaufmannChampionlevelDroneRacing2023}, production scheduling \cite{wangVirtualEntityDigital2024,hubbsDeepReinforcementLearning2020}, process design \cite{braniffEnhancedReinforcementLearningdriven2025a,reynoso-donzelliIntegratedReinforcementLearning2025a}, etc. 

With its origin in the optimal control theory, RL has also been applied for the direct control of process systems. RL has been shown to be an effective control algorithm for bioprocessing \cite{petsagkourakisReinforcementLearningBatch2020a}, distillation columns \citep{spielbergSelfdrivingProcessesDeep2019}, chemical reactors \cite{fariaOneLayerRealTimeOptimization2023}, batch processing \cite{joshiTwinActorTwin2021}, and multi-tank systems \cite{dogruOnlineReinforcementLearning2021}. Despite these advances, RL is yet to be widely adopted for the control of chemical and energy systems. The main barriers include the inherent distrust in the exploration phase of learning and the overall lack of interpretability \cite{wangTutorialReviewPolicy2025a,braniffRealtimeProcessSafety2025a,nianReviewReinforcementLearning2020a}. In RL, the exploration phase allows an agent to discover new actions that could lead to more optimal behavior, however this can be unsafe since it typically involves trying random and untested actions which could lead to undesirable or unsafe behavior. These issues prevent the RL-based control algorithms from practical implementations, especially in safety-critical process systems which require confidence in a controller’s ability to maintain safe and
stable operations \cite{fariaWhereReinforcementLearning2022,yooReinforcementLearningBatch2021a}. 

To address these challenges, extensive efforts have been made in recent years. One class of strategies is to pre-train a RL policy network using data generated from other more trusted control approaches, such as Model Predictive Control (MPC) \cite{hassanpourPracticallyImplementableReinforcement2024b,hassanpourPracticallyImplementableReinforcement2024a}. The RL agent then builds on this policy network to directly compute the control actions. If it is desired to maintain the premise of model-free RL, a linear MPC can be adopted since linear system models can easily be approximated through a variety of techniques (e.g., system identification) \cite{hassanpourPracticalReinforcementLearning2025}. If a reliable high-fidelity system model is available, model-based RL approaches can be used instead. Many of the model-based RL approaches can also provide some sort of confidence regarding safety, stability, or both together \cite{kimModelbasedSafeReinforcement2024,kimSafeModelbasedReinforcement2022,kimModelbasedReinforcementLearning2020,berkenkampSafeModelbasedReinforcement2017}. Another class of strategies use RL in a more indirect way as a supervisory role to PID control \cite{bloorControlInformedReinforcementLearning2025,chowdhuryEntropymaximizingTD3basedReinforcement2023,dogruReinforcementLearningApproach2022}. This is a promising approach to improving PID-based control but suffers from the same limitations since each sub-level controller is constrained to a single-input-single-output (SISO) control law \cite{beahrDevelopmentAlgorithmsAugmenting2024,lawrenceDeepReinforcementLearning2022}. RL has also been used for the tuning of advanced (economic) MPC \cite{grosDataDrivenEconomicNMPC2020}. In this context, the RL agent again acts in a supervisory manner by not directly computing the control actions but instead guiding controllers with guaranteed theoretical properties to make better decisions based on system feedback data \cite{alhazmiReinforcementLearningbasedEconomic2022}. An interesting example of a combined RL and MPC approach is the AC4MPC algorithm which leverages the suggested RL-based control actions as a warm-start for an MPC problem while the RL-based value function provides a better estimated terminal cost for the problem \cite{reiterAC4MPCActorCriticReinforcement2024}. Many other works have studied various ways to leverage the benefits of both MPC and RL simultaneously \cite{hedrickReinforcementLearningOnline2022,kimModelbasedReinforcementLearning2021}. 

For the development of safer RL algorithms, Lyapounov-based approaches have gained significant interest. In these algorithms, an approximate Lyapunov function is learned online. Control actions determined by the RL agent are generated in order to satisfy certain stability or safety conditions using the Lyapunov surrogate \cite{changStabilizingNeuralControl2021,chowLyapunovbasedApproachSafe2018}. This has also been extended to control Lyapunov barrier functions (CLBFs) and stochastic CLBFs which can simultaneously provide both safety and stability properties \cite{zhuReinforcementLearningOptimal2025,wangControlLyapunovbarrierFunctionbased2024a}. Other forms of safe RL include: safe exploration for linear systems \cite{marviReinforcementLearningSafety2022}, recovering safety guarantees after an offline training phase \cite{thananjeyanRecoveryRLSafe2021}, integrating principles of linear robust MPC \cite{zanonSafeReinforcementLearning2021a}, control invariant sets (CIS) \cite{wangSafeTransferReinforcementLearningBasedOptimal2024,boControlInvariantSet2023}, and Gaussian process models with chance constraints \cite{mowbraySafeChanceConstrained2022}. However, these approaches often require significant prior knowledge about the system, high-fidelity system model, and/or intensive computational power. Furthermore, these approaches must go through an exploration phase when learning the system model and/or training the RL agent which may be inherently unsafe to be implemented in practice \cite{garcia2015comprehensive}. To this end, a RL-based control algorithm is essential yet currently lacking which can improve the interpretability, stability, and computational efficiency while avoiding the unsafe exploration phase to be implemented with confidence in safety-critical systems. 

\begin{figure}[t!]
    \centering
    \captionsetup{justification=centering}
    \includegraphics[width=0.85\linewidth]{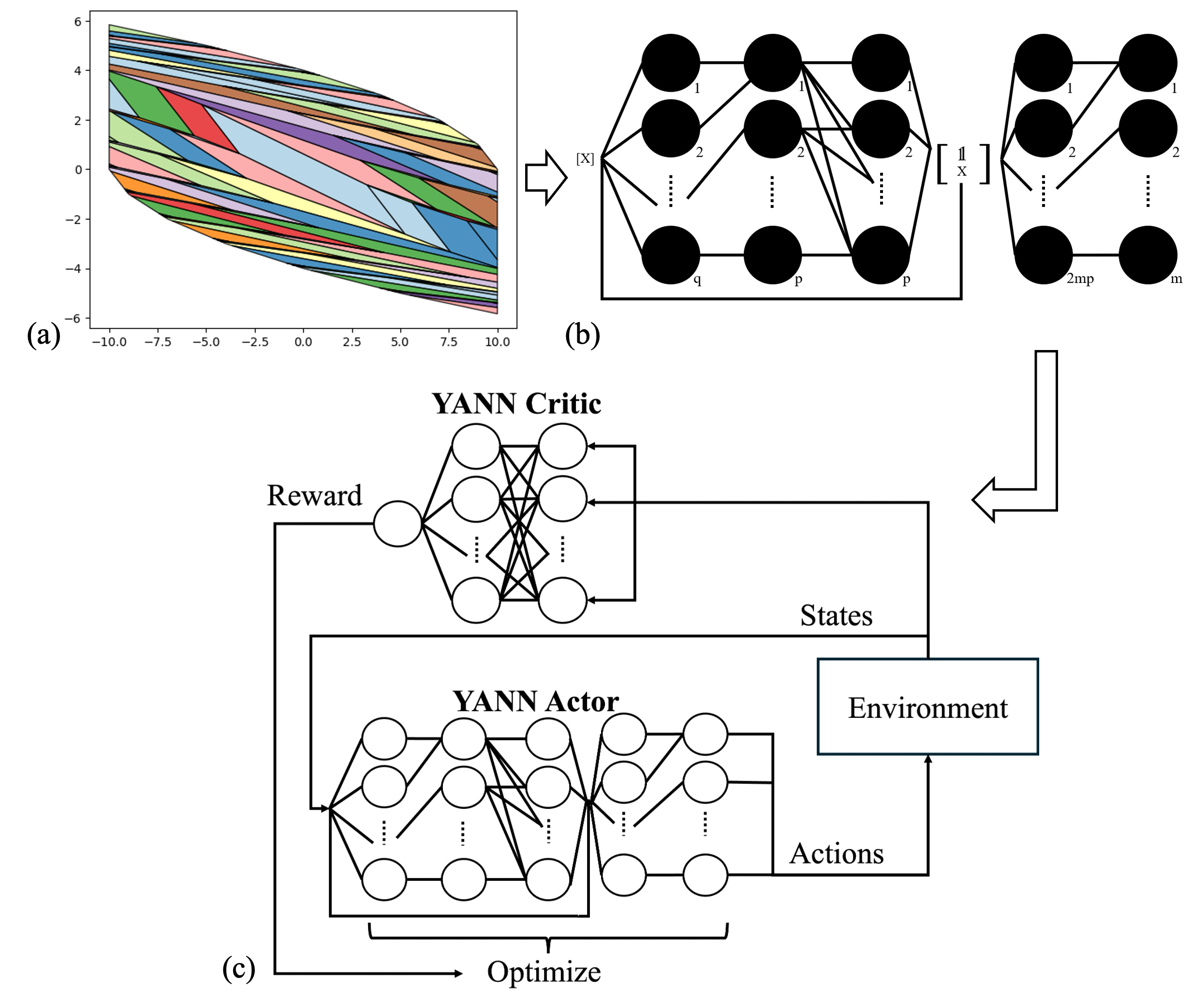}
    \caption{A schematic of the proposed RL algorithm based on YANNs. \\ (a) Multi-parametric/explicit MPC based on approximated linear model \\ (b) YANNs to exactly represent explicit control laws as piecewise linear functions \\ (c) RL initialized with YANNs for optimal nonlinear control}
    \label{fig:YANN_RL_overview}
\end{figure}

In this work, we present an RL algorithm built on the Y-wise Affine Neural Networks (YANNs), as depicted in Fig. \ref{fig:YANN_RL_overview}. Developed in our previous work \cite{braniffYANNsYwiseAffine2025a}, YANNs are a specialized neural network that can exactly represent known piecewise affine functions of arbitrary input and output dimensions defined on any amount of polytopic subdomains. Multi-parametric model predictive control (mp-MPC) presents an important application of YANNs, which theoretically computes the optimal control laws as piecewise affine functions of system states, outputs, setpoints, and disturbances \cite{pistikopoulosMultiparametricOptimizationControl2020}. Given this, the actor and the critic in an actor critic algorithm can be initialized to represent the explicit solutions and the objective to an optimal control problem via mp-MPC. Thus, YANNs can be utilized as a basis to provide an interpretable, efficient, and confident starting point for RL algorithms. This allows the algorithm to start with the full theoretical and rigorous guarantees of linear optimal control, and thereby skipping the exploration phase of RL entirely. These actor and critic networks can be created in such a way that they can approximate general nonlinear functions of arbitrary complexity using the techniques developed in this work. We further discuss how the algorithm can be continuously improved which gives confidence that the parameterized control policy will never be worse than the linear multi-parametric control policy found by solving the linear optimal control problem (OCP). The radical improvement in the control of safety-critical systems where safety constraint adherence is crucial is highlighted in a case study of a chemical reactive process. 

The remaining sections of this paper are organized as follows: Section 2 presents a brief overview of the necessary mathematical foundations and lay out the nomenclature for this work. Section 3 reviews YANNs for exact representation of piecewise affine functions and extend the network formulation to introduce nonlinearity. Section 4 introduces YANN-based RL principles including the YANN-actor, YANN-critic, and the overall RL algorithm. Section 5 demonstrates the advantages of YANN-RL through two case studies: (i) clipped pendulum, and (ii) safety-critical chemical reactor. Section 6 gives the concluding remarks and discusses future research directions.

\section{Theoretical Background}
\subsection{Introduction to reinforcement learning}
RL is a methodology that is used to solve dynamic programming problems of various kinds. It is most commonly introduced as a way to solve Markov decision processes (MDPs) which are a special class of stochastic dynamic programs. However, in this work, we apply the principles of RL with the intentions of making as few assumptions about the true system as possible while using deterministic policies. For a more thorough introduction than what follows, the readers are referred to \cite{sutton2018reinforcement,Brunton_Kutz_2022,devarakondaRecentAdvancesReinforcement2025}.

\begin{figure}[ht]
    \centering
    \includegraphics[width=0.8\linewidth]{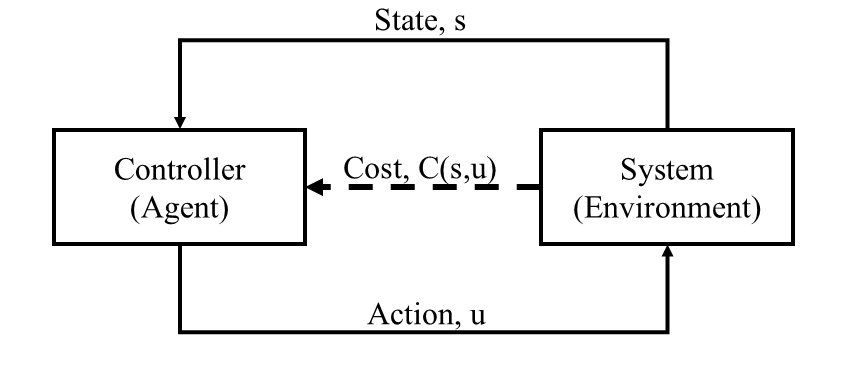}
    \caption{The terminology of RL-based control used in this work.}
    \label{fig:Basic_RL}
\end{figure}

We present the nomenclature and concepts of RL from a control-theoretic perspective. In classical control theory, a cost function is typically used to determine the optimality whereas in RL a reward function is used to this purpose. This can easily be resolved by relating cost and reward as $C=-R$, where $C$ is the cost and $R$ is the reward. Furthermore, we will use $u$ to denote an action onto the system, as is typical in controls, instead of the nomenclature $a$ which is more common in RL. In order to prevent confusion between NN inputs and states, the states of the system or environment are denoted as $s$ which is consistent with RL. A simple schematic conceptualizing the use of this terminology is presented in Fig. \ref{fig:Basic_RL}.

The central components of RL are a well-designed cost function $(C(s,u))$, the state value function $(V(s))$ or the state-action value function $(Q(s,u))$, and the policy $(\pi(s))$. The cost function is defined in such a way that optimizing control actions with respect to it yields the desired system behavior (e.g., quadratically penalizing the distance between the system state and the origin for regulation). The value function can be interpreted as a representation of how good it is for the system to exist in a particular state. It is also referred to as the cost-to-go function, which aims to provide the infinite time horizon summed cost (usually discounted) following an optimal system trajectory. For a deterministic system and policy, this is represented by Eq. \ref{eq:Value_fun} where $s_t$ is the system state at time $t$, $t_0$ is the initial time, $u_t^\star$ is the optimal action for state $s_t$ and is found by following an optimal policy $(\pi^\star(s_t))$, and $\gamma$ is a discount factor used to ensure the convergence of the function in infinite time.
\begin{equation}
\label{eq:Value_fun}
    V(s_0) = \sum_{t=t_0}^\infty \gamma^t C(s_t, u_t^\star)
\end{equation}
The state-action value function, sometimes referred to as the quality function, is heavily related to the state value function. It provides slightly more insight as it aims to give a description of how good it is to take a particular action while being in a certain state and then follow the optimal policy thereafter. This relationship is given by Eq. \ref{eq:Q_fun} where $u_0$ is the action taken on the system at time $t=t_0$.
\begin{equation}
\label{eq:Q_fun}
    Q(s_0,u_0) = C(s_0,u_0) + \sum_{t=t_1}^\infty \gamma^t C(s_t, u_t^\star)
\end{equation}
From this, Bellman's recursive optimality principle can be used to define optimality conditions for these functions when minimizing cost, as given in Eqs. \ref{eq:V_opt}-\ref{eq:Q_opt}.
\begin{equation}
    \label{eq:V_opt}
    V^\star(s_t) = \min_{u_{t}} C(s_t,u_t) + \gamma V^\star(s_{t+1})
\end{equation}
\begin{equation}
    \label{eq:Q_opt}
    Q^\star(s_t,u_t) = C(s_t,u_t) + \min_{u_{t+1}}\gamma Q^\star(s_{t+1},u_{t+1})
\end{equation}
These optimality conditions form the basis that modern RL algorithms are developed from. Typically, a NN is used to approximate one of these functions parametrically and is updated with the goal to satisfy the optimality conditions based on feedback from the system. 

\subsection{Linear optimal control}
This work will use the theoretical foundations of three types of linear optimal control: linear quadratic regulator (LQR), MPC, and mp-MPC. The OCP for LQR is defined by Eq. \ref{eq:LQR}
\begin{align}
    \label{eq:LQR}
    \min_u\quad& \sum_{t_1}^\infty\left( s_t^T Q s_t + u_t^T R u_t \right)\\
    \text{s.t.} \quad & s_{t+1} = A s_t + B u_t, \quad \forall t \in \{1, 2, \cdots, \infty\} \nonumber
\end{align}
where $s_t \in \mathbb{R}^n$ is the system state at time step $t$, $u_t \in \mathbb{R}^m$ is the control input at time step $t$, $A$ and $B$ are system matrices, $Q$ and $R$ are symmetric positive semi-definite and positive definite weighting matrices. This is a unique problem in optimal control theory since both the optimal control policy and the optimal state value function can be easily derived. Using the Bellman optimality equation (Eq. \ref{eq:V_opt}) along with the assumption that the value function takes the form $V(s) = s^TPs$, the optimal control law is derived as a function of the state value function and the parameters in the OCP as seen in Eq. \ref{eq:LQR_policy1}. 
\begin{equation}
    \label{eq:LQR_policy1}
    u_t = -(R+B^TPB)^{-1}BPA s_t
\end{equation}
Substituting $u_t$ with Eq. \ref{eq:LQR_policy1} in the Bellman equation, the discrete-time algebraic Ricatti equation (DARE) can be derived as Eq. \ref{eq:DARE}. 
\begin{equation}
    \label{eq:DARE}
    P=Q+A^TPA-A^TPB(R+B^TPB)^{-1}B^TPA
\end{equation}
This equation can be solved using various techniques (e.g., Schur method). Once a solution for the matrix $P$ is found both the optimal control law and optimal value function can be realized. 

The limitation of LQR is that these solutions are only easily found for the unconstrained system. However, in practice it is often necessary to include constraints in the decision-making process (e.g., constraints on control inputs). MPC provides an instrumental tool to handle these constraints, but it typically considers finite time. The MPC problem for linear system regulation is defined by Eq. \ref{eq:MPC}. 
\begin{align}
\label{eq:MPC}
\min_u \quad & \sum_{t=0}^{N-1} \left( s_t^T Q_w s_t + u_t^T R u_t \right) + s_N^T P s_N \\
\text{s.t.} \quad & s_{t+1} = A s_t + B u_t, \quad \forall t \in \{0,1, 2, \cdots, N-1\} \nonumber \\
& s_t \in \mathcal{S}, u_t \in \mathcal{U}, \quad \forall t \in \{0,1, 2, \cdots, N-1\} \nonumber \\
& s_{N-1} \in \mathcal{S}_f \nonumber
\end{align}
where $s_t \in \mathbb{R}^n$ is the system state at time step $t$, $u_t \in \mathbb{R}^m$ is the control input at time step $t$, $A$ and $B$ are system matrices, $Q_w$ and $R$ are symmetric positive semi-definite and positive definite weighting matrices, $P$ is the solution to DARE (Eq. \ref{eq:DARE}) to give an approximate infinite horizon cost-to-go, $N \in \mathbb{N}$ is the prediction horizon, $\mathcal{S} = \{ s \in \mathbb{R}^n : F_s s \leq g_s \}$ is the state path constraint polytope, $\mathcal{U} = \{ u \in \mathbb{R}^m : F_u u \leq g_u \}$ is the input constraint polytope, and $\mathcal{S}_f = \{ s \in \mathbb{R}^n : F_f s \leq g_f \}$ is the terminal set which is a control invariant polytope used for recursive feasibility. 

This problem can be solved online with rolling horizon using quadratic programming (QP) algorithms. An alternative approach, which is utilized in this work, is to equivalently reformulate this MPC problem into mp-MPC as given in Eq. \ref{eq:mpMPC}, to solve using multi-parametric quadratic programming (mp-QP) algorithms. 
\begin{align}
\label{eq:mpMPC}
J^\ast(\theta) =\min_u \quad & u^THu +u^TZ\theta+\theta^T\hat{M}\theta \\
\text{s.t.} \quad & Gu\leq S\theta +W \nonumber \\
& CR_A\theta\leq CR_b \nonumber
\end{align}
where $\theta$ is the parametric set comprising the state $s$ at the current time step $t=1$. $u$ is the vector of $u_t,\text{ }\forall k\in[1,2,\cdots,N]$ with $N$ being the prediction horizon from Eq. \ref{eq:MPC}. Constraint matrices $G$, $S$, and $CR_A$, and constraint vectors $W$ and $CR_b$ are derived from the system matrices and polytopic constraints in Eq. \ref{eq:MPC}. Weighting matrices $H$, $Z$, and $\hat{M}$ are derived from the system matrices and MPC weighting matrices in Eq. \ref{eq:MPC}. Similar approaches to reformulate other types of MPC problems such a setpoint tracking or disturbance rejection can be used which include necessary additional parameters in the $\theta$ vector (e.g., setpoints) \cite{pistikopoulosMultiparametricOptimizationControl2020,tian2021simultaneous}.

The advantage for mp-MPC is that this problem can be solved offline to give the optimal control law as a piecewise affine function that inherits all mathematical properties of the original formulation such as recursive feasibility and stability guarantees for the linear system. This offline solution can also enable faster control since a simpler function needs to be evaluated instead of solving a dynamic optimization problem in real time. The domain of this function is made up of closed connected but non-overlapping convex polytopes that are each a subset of the parametric constraints in Eq. \ref{eq:mpMPC}. This optimal control law representation is given in Eq. \ref{eq:PWAF}. 

\begin{equation}\label{eq:PWAF}
u^\ast(\theta) =\begin{cases}
        K_1\theta + r_1,  \theta \in CR^1 = \{ CR_A^1\theta \leq CR_b^1\} \\
        \quad \vdots \\
        K_p\theta + r_p,  \theta \in CR^p = \{ CR_A^p\theta \leq CR_b^p\} \\
         \end{cases}
\end{equation}
where $K_i$ and $r_i$ are the coefficient matrix and constant vector for the explicit solution defined on polytope $i$, and $\{CR_A^i\theta \leq CR_b^i\}$ defines the $i^{th}$ polytope $CR^i$ (i.e., critical region). An important property of this function is that it is continuous. That is, the solution to the optimal control problem defined on adjacent polytopic subdomains are equivalent at their boundary.

\section{Y-wise Affine Neural Networks and Nonlinear Extensions}
\subsection{An overview of YANNs}
\begin{figure}[t]
    \centering
    \includegraphics[width=0.8\linewidth]{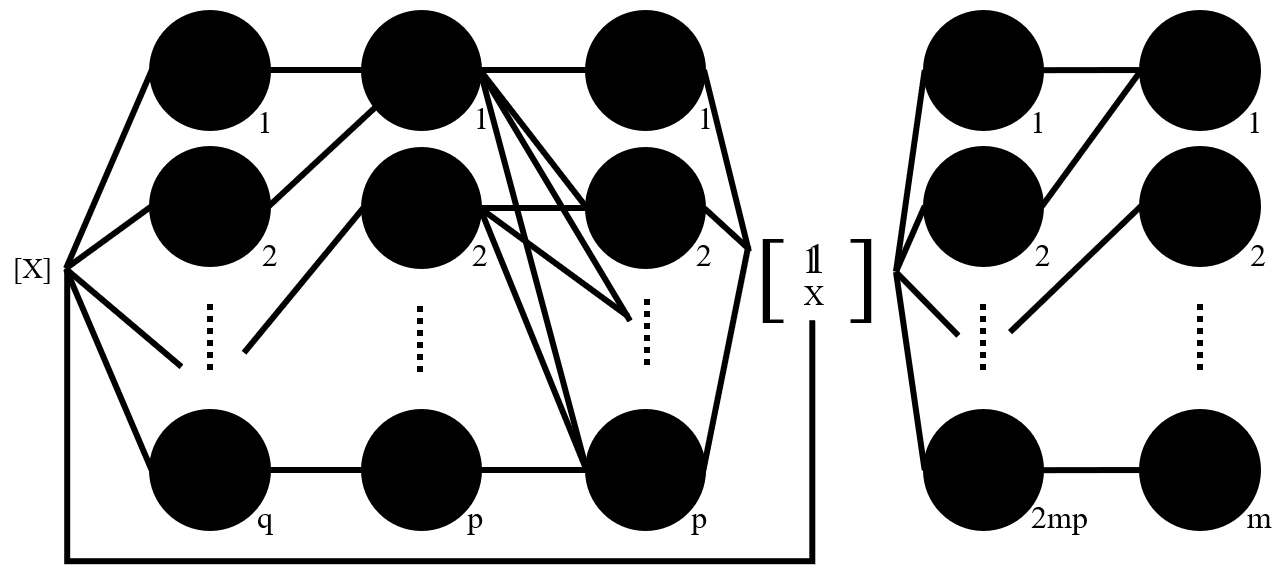}
    \caption{Y-wise Affine Neural Network architecture. }
    \label{fig:YANN}
\end{figure}

YANNs are a specific architecture that we have developed in our prior work \cite{braniffYANNsYwiseAffine2025a}. They are capable of exactly representing piecewise affine functions with any dimensional inputs and outputs defined on any number of polytopic subdomains. The YANN is a 5-layer neural network that exactly reformulates these known functions into a continuously exact representation across the full continuous domain space. YANNs can be leveraged to represent the explicit control solutions obtained from mp-MPC (Eq. \ref{eq:PWAF}). In this case, YANNs become a NN-based controller that inherits the theoretical guarantee of stability and recursive feasibility from mp-MPC for linear systems. This features a step change from previous approaches that require additional methods to provide resemblance of these essential control-theoretic properties. A YANN is represented by Fig. \ref{fig:YANN}. The first three layers determine the active subdomain for a given input, i.e. to check $\theta \in \{CR_A^i\theta \leq CR_b^i\}$. The last two layers evaluate the appropriate subfunction, i.e. to compute the corresponding $u^\ast(\theta) = K_i\theta + r_i$. $q$ is the total amount of inequalities governing the polytopic subdomains in their half-space representation, $p$ is the number of polytopic subdomains (or critical regions), $m$ is the number of outputs for the function, $\mathbbm{1}$ represents the solution to an indicator function defined by a polytopic subdomain, and the input $X$ is the set of parametric variables (e.g., states at the current time step for regulation control). A formal proof of YANNs can be found in Braniff and Tian \cite{braniffYANNsYwiseAffine2025a} which explains how to determine the specific weights and biases. 

YANNs thus provide an attractive tool for RL by being able to initialize policy networks with a control law that has theoretical confidence in its solutions, at least with respect to a linearly approximated model. The second advantage to using YANNs is that, by leveraging inference speed-ups for neural networks, they can evaluate these functions much faster than existing techniques. Two computational examples have been presented in \cite{braniffYANNsYwiseAffine2025a} comparing YANNs with the conventional mp-MPC function evaluation via Python Parametric OPtimization Toolbox (PPOPT) \cite{kenefake2022ppopt}: (i) for explicit control defined by 9 critical regions, YANN evaluates seven times faster, (ii) for explicit control defined by over 2000 critical regions, YANN decreases the inference time by more than 30\%.

\subsection{Introducing nonlinearity}
Despite the above advantages, the YANN is limited in its ability to learn more complex functions or to be trained from data at all. The initial layers pose a challenge by not being able to be updated via gradient-based approaches (e.g., backpropagation) since they use the binary step function as activation function. The subfunction evaluation layers (layers 4 and 5) are also limited to linear expressibility since they only evaluate the corresponding linear control law. To this end, it is necessary to add nonlinear expressibility to the YANN before it can be used in an algorithm like RL since nonlinear systems may require a more complex control law than the form of piecewise affine function. However, it is essential to retain the mathematical results from mp-MPC to better initialize the network and to deploy the NN-based controller with confidence at the beginning before further training. To enforce this, we add network layers that are initially evaluated to zero for any input. In a straightforward way, this can be accomplished by assigning all weights and biases to 0. But this is not useful as these parameters may not be able to be updated or if they are updated they would all update uniformly, therefore not allowing general nonlinear expressibility. In the proofs that follow, we show how NNs can be formulated with this desired property to retain the initial known mathematics of the YANN, while being trainable for nonlinear expressibility. 

\subsubsection{One-layer neural networks} 
The problem is to initialize a single layer neural network of arbitrary size such that it will provide a solution of zero for all inputs and still be trainable using modern parameter updating techniques. 
\begin{thm} 
\itshape
\label{thm:one-layer}
    A single-layer neural network with any dimensional input and output that has an even amount of nodes can be initialized with randomly generated weights and biases such that the output is always a vector of zeros for any and all possible inputs if the activation function is defined as zero at an input of zero. 
\end{thm}
\begin{pf}
\upshape
    Consider a single-layer neural network with input $X = [x_1,x_2,...,x_n] \in \mathbb{R}^n$ and output $Y\in \mathbb{R}^1$. It has an activation function defined such that $\sigma(0) = 0$. Split the nodes into two evenly sized groups. Using some random weight and bias assigning technique, generate weights and biases of the first group of nodes and define them as $W_h$ and $B_h$ respectively. Assign the weights and bias of the second group of nodes as $-W_h$ and $-B_h$. Solving for $Y$ gives:
    \begin{align*}
        Y &= \sigma(WX+B)\\
        &=\sigma(W_hX+B_h-W_hX-B_h)\\
        &=\sigma([0]) = [0]\\
        &\therefore Y=[0]
    \end{align*}
\end{pf}
\begin{cor}
\itshape
    A single-layer neural network with any dimensional input and output that has an odd amount of nodes can be initialized with randomly generated weights and biases such that the output is always a vector of zeros for any and all possible inputs if the activation function is defined as zero at an input of zero.
\end{cor}
\begin{pf}
\upshape
    Consider a single-layer neural network with input $X = [x_1,x_2,...,x_n] \in \mathbb{R}^n$ and output $Y\in \mathbb{R}^1$. It has an activation function defined such that $\sigma(0) = 0$. Let there be $2h+1$ amount of nodes in the network. Split the nodes into three groups, two evenly sized groups of size $h$ and a single node. Using some random weight and bias assigning technique, generate weights and biases of the first group of $h$ nodes and define them as $W_h$ and $B_h$ respectively. Assign the weights and bias of the second group of $h$ nodes as $-W_h$ and $-B_h$. Assign the single node to have a weight of $W_{single}=[0]$ and a bias of $B_{single}=0$. Solving for $Y$ gives:
    \begin{align*}
        Y &= \sigma(WX+B)\\
        &=\sigma(W_hX+B_h-W_hX-B_h+[0]X+0)\\
        &=\sigma([0]) = [0]\\
        &\therefore Y=[0]
    \end{align*}
\end{pf}
This symmetric weighting may give the impression that the network is not generally expressible. However, the symmetry is broken after the first parameter update assuming that a gradient-based update rule is used and that the activation function has a non-zero gradient at zero, $\frac{\partial\sigma}{\partial x}|_{x=0}\neq 0$. For example, assume that the network underestimates the true target and that all inputs to the network are positive. All weights and biases will increase. Namely, the positive parameters will become more positive and the negative parameters will take a step towards the positive direction. This breaks the symmetry since parameters that had the same magnitude, but opposite signs before the update will have different magnitudes and could have different signs. In this way, the NN would retain the potential expressibility of any other network of the same size and architecture.

\subsubsection{Two-layer neural networks} 
The problem is to initialize a two-layer neural network of arbitrary size such that it will provide a solution of zero for all inputs and still be trainable using modern parameter updating techniques. 

\vspace{0.1 in}
\begin{thm}
\label{thm:two-layer}
\itshape
    A two-layer neural network can be initialized to yield an output vector of all zeros for any and all inputs if the first layer is developed using the technique presented in Theorem \ref{thm:one-layer}, the second layer uses an activation function that is defined as zero at an input of zero, and the biases of the second layer are initialized as zero.
\end{thm}
\begin{pf}
\upshape
      Consider a two-layer neural network with input $X = [x_1,x_2,...,x_n] \in \mathbb{R}^n$ and output $Y^{[2]}\in \mathbb{R}^1$. The first layer is initialized by following the methodology presented in Theorem \ref{thm:one-layer}. The activation function of the second layer is defined such that $\sigma^{[2]}(0) = 0$. Let the biases in the second layer be initialized as zero, $B^{[2]} = 0$. The weights of the second layer can be defined using any parameter generating technique. From Theorem \ref{thm:one-layer}, it is known that the output of the first layer is a vector of zeros, $Y^{[1]}=[0]$, where the dimension is defined by the amount of nodes in layer two. Solving for $Y^{[2]}$ gives:
      \begin{align*}
          Y^{[2]} &= \sigma^{[2]}(W^{[2]}Y^{[1]}+B^{[2]})\\
          &=\sigma^{[2]}(W^{[2]}[0]+[0])\\
          &=\sigma^{[2]}([0])=[0]\\
          &\therefore Y^{[2]}=[0]
      \end{align*}
\end{pf}
Assuming that a gradient-based parameter update rule is used, the network will be as expressive as any other network of the same size and architecture. To examine how the expressibility is maintained, consider one gradient-based update. In the first update all weights in the second layer will be unchanged since the input to the second layer is a vector of all zeros and thus there is no gradient. The biases in this layer however may still be updated. Theorem \ref{thm:one-layer} has discussed how the weights of the first layer will break symmetry. This will give a non-zero input to the second layer after one update. During the second update the weights of the second layer may have non-zero gradients since the first layer is now passing non-zero information. After the first update, the network weights can be fully updated. Therefore, the expressibility is not compromised.

\subsubsection{Applying these methods}
These proofs provide a mathematical basis for adding nonlinear expressibilty to YANNs for generalizing to complex functions without sacrificing the known initialization. The proofs use arbitrary parameter generating techniques to provide a general strategy. However, in practice it is best to generate these parameters with small magnitudes (e.g., a maximum of $0.01$). This is to avoid any large changes when going through the training updates so that the network can provide similar outputs to the YANN. The proofs have constraints on the activation functions being used in each layer. Nevertheless, these are mild constraints. The widely used hyperbolic tangent function (tanh) meets all the necessary conditions for the first layer. Both tanh and rectified linear unit (ReLU) meet the necessary conditions for any layer other than the first layer, since the nonzero gradient at zero is no longer strictly necessary past layer one. Furthermore, the approach used in Theorem \ref{thm:two-layer} may be applicable to networks of larger depths. As long as the activation function of each layer is defined as zero at zero inputs and the first layer is created according to Theorem \ref{thm:one-layer}, the overall network will always provide a zero output initially but still be trainable since all network parameter can be fully updated at the second update step.

\section{YANN-based RL}
\subsection{YANN-actor}
The YANN-based actor network is simple to establish since the original YANN formulation can represent known optimal control laws. To develop a YANN-initialized policy network, an mp-MPC problem (Eq. \ref{eq:mpMPC}) needs to be solved offline in order to find a piecewise affine explicit control law. A simplified linear system model can be used, e.g. approximated from process data. After, the YANN can be created following the steps in our previous work \cite{braniffYANNsYwiseAffine2025a}. More nodes and/or layers can be added within the YANN according to Section 3.2, so that it can be trained to represent more complex control functions by directly interacting with the system online. 

\begin{figure}[h]
    \centering
    \includegraphics[width=0.8\linewidth]{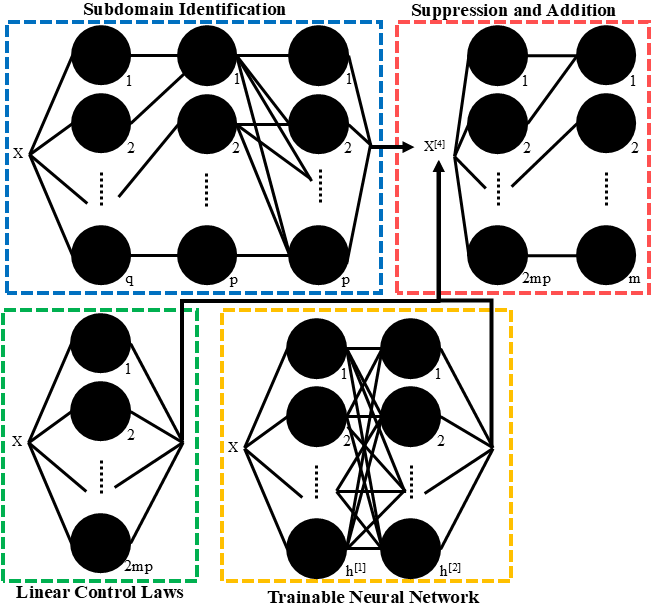}
    \caption{YANN-actor network architecture.}
    \label{fig:YANN-actor}
\end{figure}

In this work, we generate the nonlinear expressibility of the YANN-actor by injecting two-layer neural networks initialized according to Theorem \ref{thm:two-layer} in parallel with the computation of the linear control laws. One advantageous feature of the YANN is being able to locate a subset of nodes within the neural network that govern the control law for a certain partitioning of the state space (i.e., the critical region). To maintain this feature in YANN-actor, another layer must be added which acts as a suppression layer. This is better visualized in the graphic of the YANN-actor in Fig. \ref{fig:YANN-actor} where $q$ is the total amount of inequalites governing the polytopic subdomains in their halfspace representation, $p$ is the number of polytopic subdomains, $m$ is the number of manipulated variables, the input $X$ is the set of parametric variables but can be extended to other pieces of information if desired for the trainable NN component, $h^{[i]}$ is the number of nodes in the $i^{th}$ layer of the trainable NN component, and $X^{[4]}$ is the concatenated vector of the outputs between the subdomain identification, linear control law, and trainable NN components which is used as the input to the suppression and addition layers.

This structure can be interpreted in the following way. First, three vectors are computed simultaneously: (i) a vector of binaries relating to the solutions of indicator functions for the subdomains of the piecewise control law (blue box in Fig. \ref{fig:YANN-actor}), (ii) each and every piece of the control law (green box in Fig. \ref{fig:YANN-actor}), (iii) the output a two-layer NN for each piece of the control law (yellow box in Fig. \ref{fig:YANN-actor}). This information is concatenated and used to determine the solution for the active subdomain by suppressing inactive subfunction solutions to zero using ReLU bounding and then adding up the remaining values (red box in Fig. \ref{fig:YANN-actor}). This approach is adopted so that no two individual control laws interfere with each other when updating. An alternative way to conceptualize the YANN-actor is to envision that, for each polytopic subdomain that has been identified by the mp-MPC problem, the new control law is defined by a linear function plus a trainable NN. This allows each individual linear control to be uniquely updated with the updates being mutually exclusive from other control laws and also provides a degree of interpretability when examining the YANN-actor. From a mathematical standpoint, this makes sense and can allow for smaller networks to be used as opposed to one large NN that would have to represent the highly complex relationship of the true optimal control law across the full state space. Instead, each partition of the state space as identified by active constraints in the MPC problem feature their own unique control law which is consistent with optimization theory.  

\subsection{YANN-critic}
The critic in an actor critic method serves as a value function, either state or state-action, which comes from optimal control theory as discussed in Section 2. To achieve the overarching goal of deploying RL with confidence, it is necessary to derive the explicit form of these functions for linear systems. The state value function for linear unconstrained systems has been derived previously when introducing LQR as it is the solution to the DARE (Eq. \ref{eq:DARE}). However, the state-action value function is less commonly used in classic control theory. Given this, we derive it below for the LQR problem (Eq. \ref{eq:LQR}). The final solution has also been previously reported by Bradtke \cite{bradtkeReinforcementLearningApplied1992}. Let $C$ be the cost function, $C(s_t, u_t)=s_t^TQ_ws_t+u_t^TRu_t$, where $Q_w$ and $R$ are symmetric positive semi-definite and positive definite weighting matrices. Evaluating the state-action value function gives:
\begin{equation*}
    Q^\star(s_t,u_t) = C(s_t,u_t) + \min_{u_{t+1}}\gamma Q^\star(s_{t+1},u_{t+1})
\end{equation*}
\begin{equation*}
    \min_{u_{t+1}}\gamma Q^\star(s_{t+1},u_{t+1}) = \gamma V^\star(s_{t+1}) = s_{t+1}^TPs_{t+1}
\end{equation*}
\begin{equation*}
    Q^\star(s_t,u_t) = s_t^TQ_ws_t+u_t^TRu_t + \gamma s_{t+1}^TPs_{t+1}\\
\end{equation*}
\begin{equation*}
    s_{t+1} = As_t+Bu_t
\end{equation*}
\begin{align*}
    Q^\star(s_t,u_t) & = s_t^TQ_ws_t+u_t^TRu_t + \gamma (As_t+Bu_t)^TP(As_t+Bu_t)\\
    &= s_t^T(Q_w+\gamma A^TPA)s_t+2s_t^T(\gamma A^TPB)u_t+u_t^T(R+\gamma B^TPB)u_t
\end{align*}

There are two interesting things to note about this equation. First, the DARE can be similarly derived as Section 2 by noting that $Q^\star(s_t,u_t^\star) = V^\star(s_t)$ if $u_t^\star$ is the optimal action. Second, this is equivalent to the objective function of the corresponding mp-MPC problem (Eq. \ref{eq:mpMPC}) for a one-step operating horizon. This implies that for the decades of research involving mp-MPC, theorists and engineers have been using an unraveled form of the state-action value function. Furthermore, this suggests that when shifting MPC (Eq. \ref{eq:MPC}) into multi-parametric form, the objective function transforms fully into a state-action value function. To the best of our knowledge, we are the first to bring this into consideration. 

\begin{figure}[b!]
    \centering
    \includegraphics[width=0.7\linewidth]{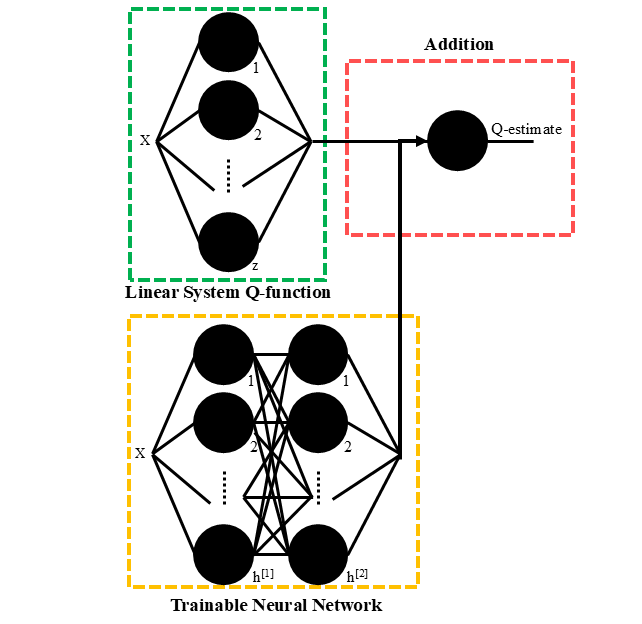}
    \caption{YANN-critic network architecture.}
    \label{fig:YANN-critic}
\end{figure}

A schematic showing the structure of the YANN-critic is provided in Fig. \ref{fig:YANN-critic}. To achieve an exact representation of this function via an NN, the resulting coefficient matrices $((Q_w+\gamma A^TPA),2(\gamma A^TPB),(R+\gamma B^TPB))$ are embedded into the weights of a linear layer using zero bias. Following this, a matrix multiplication layer is used to complete the computation of the squared and bilinear terms. A third linear layer with an all-one weight matrix and zero bias sums all of these values to yield the exact Q-value for the linear system. To add the nonlinear expressibility, a two-layer NN initialized according to Theorem \ref{thm:two-layer} is added in parallel to the previously discussed layers. The YANN-critic adds both the linear system Q-value and the output of this auxiliary NN to provide a Q-value estimate that better represents the true state-value function. The advantage to building the network in this way is that the parameters of the embedded NN and the parameters defining the linear system state-action value function can be updated simultaneously. The nomenclature of YANN-critic is meant to be consistent with that of the YANN-actor, but it is important to note that they are based on different formulations and come from different principles of control theory. 

\subsection{YANN-based DDPG} 
In this section we present and discuss a YANN-RL algorithm based on the actor-critic algorithm deep deterministic policy gradient (DDPG). In this work, DDPG is chosen as an RL algorithm that has been widely adopted in various applications and has been used towards the control of chemical and energy systems \cite{patelPracticalReinforcementLearning2023,panjapornponReinforcementLearningControl2022,siraskarReinforcementLearningControl2021,bangiDeepReinforcementLearning2021}. Additionally, it is a deterministic policy algorithm which is necessary when using a YANN-actor. The YANN-DDPG methodology is presented in Algorithm 1 which makes a few notable modifications to the original DDPG algorithm presented in Lillicrap et al. \cite{lillicrapContinuousControlDeep2019}. 

\begin{algorithm}
\label{alg:YANN-DDPG}
    \caption{YANN-DDPG}
    \begin{algorithmic}[1]
        \State Obtain a linearized model for the environment (e.g., Jacobian linearization, system identification, etc.)
        \State Use the linear model to create an MPC problem (e.g., Eq. \ref{eq:MPC})
        \State Reformulate the MPC problem into an mp-MPC problem and solve to obtain the piecewise-affine control law
        \State Use the piecewise affine control law to initialize a YANN-actor, $\pi_{\theta}(s_t)$
        \State Use the system model and MPC problem weighting matrices to initialize a YANN-critic, $Q_{\phi}(s_t,u_t)$
        \State Initialize target networks $\pi_{\theta_{target}}(s_t)$ and $Q_{\phi_{target}}(s_t,u_t)$ with $\theta_{target} \gets \theta$, $\phi_{target} \gets \phi$
        \State Initialize replay buffer, $\mathcal{R}$
        \For{episode = 1, M}
        \State Observe state $s_t$ and compute action, $u_t = \pi_\theta(s_t)$
        \State Execute action $u_t$ and observe cost, $C_t$, and new state, $s_{t+1}$
        \State Store transition $(s_t,u_t,C_t,s_{t+1})$ in $\mathcal{R}$
        \State Sample a random batch of $\mathcal{N}$ transitions $(s_i,u_i,C_i,s_{i+1})$ from $\mathcal{R}$
        \State Compute targets as:
        \begin{equation*}
            y_i = C_i+\gamma Q_{\phi_{target}}(s_{i+1},\pi_{\theta_{target}}(s_{i+1}))
        \end{equation*}
        \State Update the YANN-critic by minimizing the loss: 
        \begin{equation*}
            \mathcal{L}=\frac{1}{\mathcal{N}}\sum_i^\mathcal{N} (y-Q_\phi(s_i,u_i))^2
        \end{equation*}
        \State Update the YANN-actor using the sampled policy gradient: \begin{equation*}
            \nabla _\theta J=\frac{1}{\mathcal{N}}\sum_i^\mathcal{N} Q_\phi(s_i,\pi_\theta(s_i))
        \end{equation*}
        \State Update target networks using Polyak averaging:
        \begin{equation*}
            \theta_{target} = \tau_{actor}\theta+(1-\tau_{actor})\theta_{target}
        \end{equation*}
        \begin{equation*}
            \phi_{target} = \tau_{critic}\phi+(1-\tau_{critic})\phi_{target}
        \end{equation*}
        \EndFor
    \end{algorithmic}
\end{algorithm}

The most critical change is that the weights and biases defining the actor and the critic are not fully randomly generated. They are initialized using specific tools from control theory and with the approaches presented in Section 3. Another key adjustment is the complete and total lack of random exploration. It is a necessary component of the DDPG algorithm to add exploration noise to the computed action (e.g., Ornstein-Uhlenbeck or Gaussian) in order to search through the state space and develop a basis for the exploitation phase. The exploration is not needed in this YANN-DDPG algorithm since the policy starts with a good basis across the entire state space, and thus we can begin the algorithm already exploiting the critic. Skipping the exploration phase also allows a much safer and smoother operation since random perturbations on the system are avoided. This proposed algorithm begins with the full confidence of linear optimal control which is a significant departure from other RL algorithms. 

\subsection{Monotonic policy improvements} 
The seminal results of Khakade and Langford \cite{Langford2002} give critical insight into how policies should be updated to achieve a continuous improvement in performance. The central idea is that there exists some regions around the current policy such that updating the policy to remain  within this region will lead to a guaranteed improvement. This idea has been fundamental in several works such as trust region policy optimization (TRPO) \cite{schulmanTrustRegionPolicy2017}, safe policy iteration \cite{pirottaSafePolicyIteration2013}, and several others \cite{scherrerLocalPolicySearch2014,dalalSafeExplorationContinuous2018,papiniStochasticVarianceReducedPolicy2018,brunkeSafeLearningRobotics2022a}. While these theoretical guarantees are highly desired, they are often too difficult to achieve in practice. The common issues among these approaches is that they either need assumptions that are challenging to verify or they are computationally intractable. For example, TRPO, which is one of the more tractable approaches of those with guarantees, requires a second-order optimization algorithm to approximate the Hessian of KL divergence constraints. This is often far too expensive to be applicable, especially for real-time systems. Another issue is that most theoretical approaches require a large amount of sampling in order to provide any confidence at all. For these reasons, most modern and practical RL algorithms implement a heuristic approach to providing safe and continuously improving policy updates. A good example is proximal policy optimization (PPO) \citep{schulmanProximalPolicyOptimization2017} which is a direct successor of TRPO. In PPO, a small epsilon value is used to clip how much the policy updates, effectively implementing a TRPO-style algorithm without having to track through the expensive computations. One of the goals of the YANN-DDPG algorithm is to use these ideas in order to have heuristic confidence in continuously improving the policy. This is done in two ways: (i) small learning rates and small target network updates, and (ii) interpretable initializations of the actor and the critic. The small updates made to the networks in this algorithm ensure that they remain close to their representations prior to being updated. In other words, none of the networks are being updated drastically enough to push them outside of a known area of improvement, which is similar to the clipping in PPO. The interpretable initializations of both the actor and the critic provide a foundations for which this continuous improvement can occur. When either the critic or the actor is updated, they are expected to improve when making a small enough update and thus should improve over their initial representation. In this way, there is confidence that on average the control policy is never getting worse the than the previous control policy. This is not a full-proof guarantee as the algorithm still utilizes the structure of DDPG which holds very little theoretical confidence compared to that of TRPO or PPO. Nevertheless, this is an exceptional result since linear optimal control is effectively a lower bound of performance for YANN-DDPG. This statement is empirically validated in two case studies in Section 5.

\section{Case studies}
\subsection{Clipped pendulum}
In the first case study, we consider a simple pendulum about a fulcrum point which is a widely used benchmark system for RL-based control. It is desired to move the pendulum to an upright position by applying torque on it. For simplicity we consider a constrained initial position in order to avoid the swing-up problem which is a classically difficult problem in control. The motivation of this case study is to highlight the potential benefits of YANN-DDPG in comparison to DDPG. The governing dynamics of the pendulum system are represented in Eq. \ref{eq:pendulum}. The system states are: (i) the angle, $\theta$, measured from an upright position, and (ii) the angular velocity, $\frac{d\theta}{dt}$. 
\begin{align}
\label{eq:pendulum}
    \frac{d^2\theta}{dt^2} = \frac{3g}{2l}\sin(\theta) + \frac{3}{ml^2}u
\end{align}
where $u$ is the applied torque in $N \cdot m$, $g=10 \frac{m}{s^2}$ is the gravitational constant, $l=1$ m is the length of the pendulum, and $m=1$ kg is the mass of the pendulum. Towards the direction of developing a YANN-actor, these equations are linearized around the origin (pendulum in an upright and unmoving state) using the Jacobian method and then discretized for a sampling rate of $0.05s$ which is consistent with the OpenAI Gym simulator \cite{brockman2016openai}. In practice, for true model-free RL, system identification would be used instead. The resulting linear system is given in Eq. \ref{eq:linear_pendulum}. 
\begin{equation}
    \label{eq:linear_pendulum}
    A=\begin{bmatrix}
        1.0188& 0.0503\\ 0.7547&  1.0188
    \end{bmatrix},B=\begin{bmatrix}
        0.0038\\ 0.1509
    \end{bmatrix}
\end{equation}
This linear system is used to create an MPC problem similar to the one shown in Eq. \ref{eq:MPC} with $N=2$, $Q$ is the two-dimensional identity matrix, $R=0.001$, the state path and terminal constraints are $-2\leq\theta_t\leq2$, $-8\leq \frac{d\theta_t}{dt}\leq8$, the control input constraint is $-2\leq u_t \leq 2$, and $A$ and $B$ are from Eq. \ref{eq:linear_pendulum}. The constraints on the control input and on the states are consistent with the constraints of the pendulum system. 

The MPC problem is reformulated into an mp-QP (Eq. \ref{eq:mpMPC}) and solved via multi-parametric programming to give an explicit control law as a function of the states at time $t=0$. The piecewise affine control law has seven subdomains partitioned over the constrained state space. This function is used to formulate a YANN-actor. The YANN-actor has an additional two-layer NN with $16$ nodes in each layer using the $tanh$ activation function for every subdomain giving a total parameter count of $15,188$ with $12,880$ of those being trainable. This is less than the DDPG actor which features a two-layer NN using $256$ nodes in each layer giving $67,073$ total parameters, all of which are trainable. Furthermore the YANN-actor can be expected to evaluate faster given its sparsity over the DDPG actor which is fully dense. The YANN-critic uses the weighting and system matrices in the MPC problem along with two additional NN layers each of $64$ nodes using the $tanh$ activation function for a total of $4,358$ parameters. The DDPG critic uses the same exact structure of the DDPG actor. Both algorithms are initially tested on a set of the same random 10 episodes to benchmark their initial performance. Then each algorithm, YANN-DDPG and DDPG, is allowed to run for 50 episodes of training. After these training episodes, both algorithms are again benchmarked on the same 10 testing episodes. These results are shown in Table \ref{tbl:pendulum} and highlight the improvement each algorithm made.

\begin{table}[ht]
\centering
\caption{Total cost per test episode for the pendulum environment.}
\begin{tabular}{|c|c|c|c|c|c|}
\hline
\textbf{Episode \#} & \multicolumn{2}{c|}{\textbf{DDPG}} & \multicolumn{3}{c|}{\textbf{YANN-DDPG}} \\
\hline
  &Initial  &Final   &Initial  &Final  &Change  \\
\hline
1  &1,570.25  &3.83   &1.50  &1.51 & +0.01  \\
\hline
2  &1,586.54  &2.89  &0.55  &0.55  & 0.00  \\
\hline
3  &1,597.88  &4.33  &1.93  &1.93  & 0.00   \\
\hline
4  &1,585.44  &2.61  &0.15  &0.15  & 0.00   \\
\hline
5  &1,597.85  &3.58  &1.15  &1.15  & 0.00   \\
\hline
6  &1,552.79  &2.75  &0.35  &0.35  & 0.00   \\
\hline
7  &1,585.54  &2.91  &0.57  &0.57  & 0.00   \\
\hline
8  &1,603.19  &129.03  &1,354.73  &127.00  & -1,227.73  \\
\hline
9 &1,604.24  &128.99  & 1,348.27 &126.84  &  -1,221.43 \\
\hline
10 &1,585.99  &3.25  &0.74  &0.74  & +0.00  \\
\hline
Average &1,586.97  &28.42  &270.99   &26.08  &-244.91  \\
\hline
\end{tabular}
\label{tbl:pendulum}
\end{table}
The DDPG algorithm starts off much worse than the YANN-based algorithm. The average initial test cost for DDPG is almost $1,600$ whereas the YANN-actor has an initial average cost of $271$. This result highlights how the YANN-based approach can provide a significantly better starting point for RL and demonstrates the potential efficiency boosts. At some episode within the 50 training episodes, the DDPG agent performance would be similar to that of the initial YANN-actor which represents that the YANN-DDPG approach could save training time over the DDPG algorithm. Another interesting result is that the initial YANN-actor outperformed the trained DDPG agent on $8$ out of the $10$ test episodes while the final YANN-actor outperformed the trained DDPG agent on all test episodes. This is evidence that the YANN-based algorithm is much more efficient in the learning process since it is able to consistently outperform DDPG when both are trained for the same amount of episodes. The results confirm the expectations of continuous improvements in the policy when using small updates. The YANN-based algorithm is improved dramatically, e.g. for test episode $8$ where cost drops by over $1,200$.  Test episode $1$ is the only episode where the YANN-actor observed worse performance after training, while cost increases marginally by $0.01$. This shows that the continuous improvement is more heuristic confidence than it is a true guarantee. Nonetheless, having linear optimal control as an effective lower bound is a promising result that has been lacking in RL for control.

\subsection{Safety-critical reactor}
In the second case study, we consider a safety-critical reaction process conceptualized from a real-world process incident \cite{braniffHierarchicalMultiparametricProgramming2024b,aliDynamicRiskbasedProcess2023d}. It is a continuous stirred tank reactor (CSTR) that is represented by the equations given in Eq. \ref{eq:CSTR} where modeling and parameter information is provided in the Appendix in Table A.1. The main reactions are also given below.

   \begin{subequations}\label{eq:CSTR}
    \begin{equation}
        \frac{dC_{A}}{dt}=\frac{F_{A,in}-q_{out}}{V}C_A-k_{1}C_{A}C_{B}
    \end{equation}
    \begin{equation}
        \frac{dC_{B}}{dt}=\frac{F_{B,in}-q_{out}}{V}C_A-k_{1}C_{A}C_{B}
    \end{equation}
    \begin{equation}
        \frac{dC_{S}}{dt}=\frac{F_{C,in}-q_{out}}{V}C_A-k_{2}C_{S}
    \end{equation}
    \begin{equation}
         \frac{dT}{dt}=\frac{q_{out}}{V}(T_{in}-T)+\frac{\sum(-\Delta H_{k}r_{k})-\frac{UA_x}{V}(T-T_{c})}{\rho C_{P}}
    \end{equation}
\end{subequations}

\newpage
\noindent \underline{Reaction 1:} 
\begin{center}
Methylcyclopentadiene (A) + Sodium (B) \\ $\xrightarrow{Diglyme (S)}$ Sodium Methylcyclopentadiene (C)  + Hydrogen (D)
\end{center} 
\noindent \underline{Reaction 2:} 
\begin{center} Diglyme (S) $\xrightarrow{Sodium (B)} $ Hydrogen (D) + Byproduct
\end{center}

Hydrogen gas is a product of both reactions which causes major safety concerns. Additionally, the second reaction is highly exothermic which leads to thermal runaway at high temperatures. For these reasons, it is essential to operate this reactor below temperatures of $480$ K which is the temperature at which the rate of reaction 2 becomes significant. This defines the safety limits for the operation of the CSTR. 

For simplicity, in this control study we consider only the regulation of this safety-critical CSTR to steady state while adhering to the safety constraint of keeping the temperature below $480$ K. The dynamic equations in Eq. \ref{eq:CSTR} are linearized around the steady state using the Jacobian method and then discretized to a sampling rate of $1/60$ s. The resulting linearized system is given in Eq. \ref{eq:CSTR_linear}.
\begin{align}
\label{eq:CSTR_linear}
    &A = \begin{bmatrix}
         0.9506& 0& 0& 0\\
        -0.0484& 0.9943& 0& 0\\
        0& 0& 0.9909& 0\\
        0.6970& 0.0678& 0& 1.0030\\
    \end{bmatrix}
    B=\begin{bmatrix}
        0\\0\\0\\-0.0007
    \end{bmatrix}
\end{align}

This linear system is used to create an MPC problem similar to the one shown in Eq. \ref{eq:MPC} with $N=2$, $Q = 0.1\times I_2$, $R=0.0001$, the state path and terminal constraints in deviation variables are $-2\leq \overline{C_i}\leq2$ for each species $i$, $-70\leq \overline{T}\leq5$, the control input constraint is $-55\leq u_t \leq 55$, and $A$ and $B$ are from Eq. \ref{eq:CSTR_linear}. The upper bound on the temperature deviation corresponds to the safety-critical constraint for this system. 

Similar to the pendulum case study, the MPC problem is reformulated into an mp-QP (Eq. \ref{eq:mpMPC}) and solved via multi-parametric programming to give an explicit control law as a function of the states at time $t=0$. The piecewise affine control law again has seven subdomains partitioned over the constrained state space. This function is used to initialize a YANN-actor. The YANN-actor has an additional two-layer NN with $8$ nodes in each layer using the $tanh$ activation function for every subdomain giving a total parameter count of $5,320$ with $3,416$ of those being trainable. This is less than the DDPG actor which features a two-layer NN using $256$ nodes in each layer giving $67,329$ total parameters, all of which are trainable. The YANN-critic uses the weighting and system matrices in the MPC problem along with two additional NN layers each of $64$ nodes using the $tanh$ activation function for a total of $4,545$ parameters. The DDPG critic uses the same exact structure of the DDPG actor. Any departure in the amount of parameters needed between this case study and the previous one for the NNs in DDPG or for the YANN-critic can be explained by needing more inputs (i.e., more system states). The YANN-actor has significantly less complexity in this example to showcase its ability to still perform well with limited parameters. Both algorithms are initially tested on a set of the same random 10 episodes to benchmark their initial performance. Then each algorithm is allowed to run for 50 episodes of training. The training episode stops if the safety-critical constraint is violated ($T>480$ K) and a penalty of $1E5$ is added to the immediate cost to encourage behavior that avoids violating safety limits. The amount of time the algorithm violates this safety condition is noted during the training process. After the training episodes, both algorithms are again benchmarked on the same 10 testing episodes. For this system the stochasticity of the DDPG algorithm can play a large role in its performance. For this reason, we have included two scenarios using the same DDPG algorithm in the results, one with a good initial performance (DDPG-G) and one with a poor initial performance (DDPG-P). These results are shown in Table \ref{tbl:CSTR}.
\begin{table}[ht]
\centering
\caption{Total cost per test episode for the CSTR environment.}
\adjustbox{max width=\textwidth}{
\begin{tabular}{|c|c|c|c|c|c|c|c|}
\hline
& \multicolumn{2}{c|}{\textbf{DDPG-G}} & \multicolumn{2}{c|}{\textbf{DDPG-P}} & \multicolumn{3}{c|}{\textbf{YANN-DDPG}} \\
\hline
Safety Violations& \multicolumn{2}{c|}{1} & \multicolumn{2}{c|}{9} & \multicolumn{3}{c|}{0} \\
\hline
  Episode&Initial  &Final &Initial  &Final   &Initial  &Final  &Change  \\
\hline
1  &901.95  & 662.14       &110.99    &87.04 &30.67  &30.84 & +0.17 \\
\hline
2  &173.15  & 147.89      &1,055.08  & 634.00&58.93  &59.70 & +0.77  \\
\hline
3  &164.74  & 131.13      &976.04   &572.71 &54.77  &55.40 & +0.63   \\
\hline
4  &1,671.20  &1,357.62          &306.6     &417.88 &115.32  &113.09 & -2.23 \\
\hline
5  &193.86  & 184.65      &$>$1E5       &749.03 &75.77  &76.34 & +0.57 \\
\hline
6  &2,813.91  & 2,422.08     &842.77      &1,097.05 &297.99  &283.84 &  -14.15 \\
\hline
7  &517.99  & 334.26      &154.29    &31.11 &10.76  &10.39 &  -0.37\\
\hline
8  &905.87  & 667.46      &121.50     &96.51 &36.57  &35.86 &  -0.71\\
\hline
9 &277.34  & 301.08       &$>$1E5      &1,032.82 &110.96  &109.37 &  -1.59\\
\hline
10 &382.38  & 435.59      &$>$1E5       &1,319.69 &151.44  &147.31 &   -4.13\\
\hline
Avg. &800.24  &664.39  &$>$3E4  &603.78 &94.32  &92.22 &  -2.10\\
\hline
\end{tabular}
}
\label{tbl:CSTR}
\end{table}

The DDPG algorithm performs much worse for the CSTR system than it did for the simple pendulum example, regardless of a good random initialization (DDPG-G) with an average test cost of around $800$ or a poor initialization (DDPG-P) where the agent violated the safety constraint on three of the ten test episodes. The CSTR has more states and more restrictions on the space that can be explored due to the safety constraint which may contribute to this behavior, since the DDPG algorithm relies on exploring the state space before exploiting its structure. This may be better explained by the DDPG-P trial where violating the safety constraint numerous times leads to a better final performance. Nevertheless, in both trials the algorithm violated the safety constraint at least one time (Table \ref{tbl:CSTR}). This shows that even with an initially decent actor the algorithm may still give undesirable behavior for the real-world system. 

The YANN-DDPG algorithm performs immensely better than the simple DDPG approaches. The YANN-based RL method provides an initial performance that is around an order of magnitude better than the trained DDPG-agents. This result shows that the YANN-DDPG algorithm save more than 50 episodes worth of training time over DDPG since it cannot achieve the same performance within the training episodes. Furthermore, the training process in the YANN-DDPG trial never violated the safety constraint a single time. The heuristic confidence in expecting the YANN-agent to at worst perform as well as linear optimal control can also be seen in Table \ref{tbl:CSTR}. The YANN-based algorithm performs worse than its initialization on four of the ten episodes but only by a small margin (maximum of $0.77$) whereas it improves in the remaining six episodes up to two orders of magnitude more (maximum of $14.15$).

\section{Conclusions}
In this work we have introduced a novel RL algorithm based on the use of YANNs, which are interpretable neural networks with desirable properties for control-theoretic applications. This algorithm represents a paradigm shift in our ability to confidently deploy RL algorithms for the optimal control of chemical and energy systems. We showed how to initialize YANN-based actors and critics to exactly represent the solution and the objective to linear optimal control problems. We discussed how one can have confidence that the control policy should only improve over its linear approximation. We also showed how to initialize neural networks to always output 0 but still be trainable. We highlighted these results in two case studies where one featured the operation of a safety-critical system where the DDPG algorithm violated safety constraints numerous times while the YANN-DDPG algorithm never once did. In the future we wish to provide more rigorous theory and algorithms that can truly guarantee continuous improvement over the linear control approximation. We will also look to integrate YANNs into other well-established RL algorithms and to develop a new algorithm altogether that exploits additional properties of YANNs.

\section*{Acknowledgments}
The authors acknowledge financial support from NSF RETRO Project CBET-2312457, NSF GRFP 2024370240, and Department of Chemical and Biomedical Engineering at West Virginia University.

\newpage
\appendix
\section{Safety-critical CSTR modeling information}
\label{app1}
\begin{table}[h!]
  \centering
  \caption*{Table A.1: Defining CSTR modeling parameters and values.}
  \label{tbl:CSTR_modeling}
  \begin{adjustbox}{max width =\textwidth}
  \begin{tabular}{|l|l|}
  \hline
  State variables & $C_A, C_B, C_S$: Concentrations\\& $T$: Temperature \\ \hline
  Manipulated variable & $U$: Heat transfer coefficient \\ \hline
  Control variable & $T$: Temperature \\ \hline
  Modeling variables & $V$: Volume (4000 L)\\ &$\rho$: Mixture density (36 mol/L) \\ 
   & $C_p$: Specific heat (430.91 J/mol$\cdot$K) \\ 
  & $A_x$: Heat transfer area (5.3 m\textsuperscript{2})\\& $T_c$: Coolant temperature(373K) \\
  & $\Delta H_{k}$: (-45.6 kJ/mol, -320 kJ/mol)\\ 
  & $k_i = A_i\mbox{exp}(-\frac{E_i}{RT})$\\
  & $A_i$:($A_1 = 4\times10^{14}$, $A_2 = 1\times10^{84}$) \\
  & $E_i$:($E_1 = 1.28\times10^{5}$, $E_2 = 8\times10^{5}$ J/mol$\cdot$K) \\
  \hline
  \end{tabular} 
  \end{adjustbox}
\end{table}
\newpage

%\bibliographystyle{elsarticle-num} 
%\bibliography{bibliography}

\end{document}